\documentclass[twocolumn]{aastex631}

\usepackage{graphicx}
\usepackage{amsmath,amssymb}
\usepackage{braket}
\usepackage{subfigure}
\usepackage{float}
\usepackage{xcolor}
\usepackage{soul}
\usepackage{rotating}
\usepackage{multirow}
\usepackage{mathtools}
\usepackage{verbatim}

\usepackage{hyperref}
\hypersetup{
	colorlinks=true,
	linkcolor=red,
	citecolor=blue,
}

\newcommand{\dHy}{{\tt dHybridR}}

\newcommand{\C}{{}_{\rm cr}}
\newcommand{\B}{{}_{\rm bst}}
\newcommand{\I}{{}_{\rm iso}}
\newcommand{\D}{{}_{\rm d}}

\begin{document}

\title{Modeling the Saturation of the Bell Instability using Hybrid Simulations}

\author[0000-0002-2890-6758]{Georgios Zacharegkas}
\affiliation{High-Energy Physics Division, Argonne National Laboratory, Argonne, IL 60439, USA}
\affiliation{Kavli Institute for Cosmological Physics, University of Chicago, Chicago, IL 60637, USA}

\author[0000-0003-0939-8775]{Damiano Caprioli}
\affiliation{Department of Astronomy \& Astrophysics, University of Chicago, Chicago, IL 60637, USA}
\affiliation{Enrico Fermi Institute, The University of Chicago, Chicago, IL 60637, USA}

\author[0000-0002-2160-7288]{Colby Haggerty}
\affiliation{Institute for Astronomy, University of Hawaii, Honolulu, HI 96822, USA}

\author[0000-0002-1030-8012]{Siddhartha Gupta}
\affiliation{Department of Astronomy \& Astrophysics, University of Chicago, Chicago, IL 60637, USA}
\affiliation{Department of Astrophysical Sciences, Princeton University, 4 Ivy Ln., Princeton, NJ 08544, USA}

\author[0000-0002-4273-9896]{Benedikt Schroer}
\affiliation{Department of Astronomy \& Astrophysics, University of Chicago, Chicago, IL 60637, USA}

\begin{abstract}
The nonresonant streaming instability (Bell instability) plays a pivotal role in the acceleration and confinement of cosmic rays (CRs); 
yet, the exact mechanism responsible for its saturation and the magnitude of the final amplified magnetic field have not been assessed from first-principles. 
Using a survey of hybrid simulations (with kinetic ions and fluid electrons), we study the evolution of the Bell instability as a function of the parameters of the CR population. 
We find that, at saturation, the magnetic pressure in the amplified field is comparable with the initial CR anisotropic pressure, rather than with the CR energy flux as previously argued.
These results provide a predictive prescription for the total magnetic field amplification expected in the many astrophysical environments where the Bell instability is important. 
\end{abstract}


\section{Introduction}\label{sec:intro}

Collisionless shock waves associated with supernova remnants (SNRs) are believed to be the primary source of Galactic cosmic rays (CRs) (up to ``knee" rigidities of $\sim 10^{15}$ V), through the Diffusive Shock Acceleration (DSA) mechanism \citep{bell78a,blandford+78}. 
However, for efficient CR acceleration through DSA, CRs must be confined close to the shock, which requires the presence of very strong, turbulent magnetic fields \citep{lagage+83a,blasi+07}. 
Strong magnetic turbulence and CR acceleration are thus closely related and the study of the growth and saturation of such strong magnetic fields is crucial to explain the origin of high-energy CRs.

\citet{winske+84} originally found that in systems with a sufficiently strong CR current, right-handed modes with wavelengths significantly smaller than the CR gyroradius could be excited; 
these modes differ from the linearly polarized ones driven by the resonant CR streaming instability, which is caused by CRs in gyroresonance with Alfv\'{e}nic modes \citep{kulsrud+69,zweibel79, achterberg83}. 
Historically, only the resonant modes were believed to be important because of the gyroresonance condition for CR scattering; the non-resonant branch was ignored until \citet{bell04} showed that the right-handed modes grow faster than left-handed ones and ---most importantly--- that they can grow to non-linear levels, eventually saturating at large amplitudes.
This claim was motivated by simulations that couple a magneto-hydrodynamical (MHD) description of the thermal plasma with a kinetic treatment of the CR current \citep[][]{lucek+00,bell+01}. 
The \emph{nonresonant streaming instability}, often referred to as the ``Bell instability", occurs in many space/astrophysical environments, and in particular is crucial for the production of high-energy CRs in SNRs, as attested by global kinetic simulations of non-relativistic shocks  \citep[e.g.,][]{caprioli+13,caprioli+14a,caprioli+14b,caprioli+14c,crumley+19,marcowith+21}, as well as for the dynamics of the shocks themselves 
\citep[e.g.,][]{haggerty+20,caprioli+20}.

\subsection{Linear theory and modifications}
The linear theory for the Bell instability shows that the instability is expected to grow faster than the resonant instability when the maximally unstable wavelength, of wavenumber $k_{\max}$, is much smaller than the CR gyroradius $r_L$, i.e., $k_{\rm max} r_L\gg 1$; 
we comment on such condition extensively in what follows.

In this regime right-handed circularly-polarized modes are driven unstable and grow faster than their resonant (left-handed) counterparts, as shown, e.g., in \S4.3 of \citet{bell04} and \S3 of \citet{amato+09}.
The wavenumber of the fastest-growing mode is 
\begin{equation}\label{eqth:k_max Bell}
	k_{\max} =  \frac{2\pi}{c} \frac{J\C}{B_0} = \frac{1}{2} \left( \frac{n\C}{n_{\rm g}} \right) \left( \frac{v\D }{v_{\rm A,0}} \right) d_i^{-1} \; 
\end{equation}
and its corresponding growth rate reads
\begin{equation}\label{eqth:gamma_max Bell}
	\gamma_{\max} = k_{\max} v_{\rm A,0} = \frac{1}{2} \left( \frac{n\C}{n_{\rm g}} \right) \left( \frac{v\D }{v_{\rm A,0}} \right)\Omega_{ci} \; ,
\end{equation}
where $e$ and $m$ are the proton charge and mass, $J\C=e n\C v\D $ is the CR current, $n\C$ and $v\D $ are the CR number density and drift velocity relative to the background plasma, $B_0$ and $n_{\rm g}$ are the background magnetic field and plasma number density, $v_{\rm A,0} \equiv B_0/\sqrt{4\pi m n_{\rm g}}$ is the initial Alfv\'{e}n speed, $\Omega_{ci} \equiv eB_0/(mc)$ is the ion gyrofrequency, and $d_i \equiv v_{\rm A,0}/\Omega_{ci}$ is the ion inertial length. 

It is important to point out that a current of energetic particles may also drive a plethora of additional modes, both parallel and transverse to the magnetic field \citep[e.g.,][]{bykov+11a,malovichko+15},  as discussed in the thorough review by \citet{bret09}, where the growth rate of Weibel, two-stream, Buneman, filamentation, Bell, and cyclotron instabilities are compared for different CR parameters.
While these instabilities may be important in some environments, especially at scales smaller than the ion skin depth, global kinetic simulations of strong shocks confirm that Bell amplification is the most prominent way of producing the turbulence responsible for the acceleration of CRs to higher and higher energies \citep[e.g.,][]{caprioli+14a,caprioli+14b, caprioli+18, haggerty+19a, marcowith+21}.

The original derivation of the growth rate of the Bell instability did not account for situations where the background plasma is not strictly cold or collisionless.
The Bell instability is expected to be modified when the background plasma is sufficiently warm; 
in this regime (dubbed ``WICE'' for warm ions/cold electrons) the fastest growing wavenumber shifts to $k_{\rm wice}<k_{\max}$ and the growth rate is suppressed \citep{reville+08a,zweibel+10,marret+21}.
Additionally, the instability is modified in systems where the collisional time scale becomes comparable to the growth rate \citep{reville+07}. 
Systems with proton-neutral collisions were found to have reduced growth rates and saturated magnetic field amplitudes, while systems with proton-proton collisions resulted in a larger saturated magnetic field, owing to the suppression of temperature anisotropy generation \citep{marret+22}.
While these considerations are potentially important to select systems, they are not included in the present work.

\subsection{Simulating the Bell instability}
MHD simulations have shown that, for a fixed CR current, large amplification factors can be achieved \citep[e.g.,][]{bell04,bell05,zirakashvili+08c, matthews+17}.
Nevertheless, these simulations cannot self-consistently capture the back-reaction of the growing modes on the CRs, hence they cannot be used to assess the saturation of the Bell instability.

Using particle-in-cell (PIC) simulations, \cite{niemiec+08} found a much lower level of amplitude of the saturated magnetic field, but they also found a growth rate for the fastest growing mode which was smaller than what \cite{bell04} had predicted, putting into question the existence of the Bell instability beyond the MHD limit. 
However, PIC simulation performed by \citet{riquelme+09} showed that for the exceedingly strong currents adopted in the work of \cite{niemiec+08} a transverse  filamentary mode can grow faster than the Bell instability; 
they also found that for typical CR currents the Bell instability grows as expected and saturates to levels of $\delta B/B_0 \gtrsim 10$. 
The saturation was found to be caused by the background plasma being accelerated in the direction of the CR drift velocity, which reduces the CR current $\mathbf{J}\C$ that drives the instability. 
Similar results have been found in the PIC simulations performed by other groups as well \citep{stroman+09,ohira+09b, gupta+21}.

\cite{gargate+10,weidl+19a,weidl+19b} used hybrid simulations (with kinetic ions and fluid electrons) to follow the instability on longer time-scales, well into the non-linear regime when saturation occurs. 
Their results supported a saturation mechanism similar to that described by \cite{riquelme+09}, who used full PIC simulations: saturation occurs due to the deceleration of CRs and simultaneous acceleration of the background plasma, which reduces the CR current, an effect also seen in the earlier work of \citet{lucek+00}.

The vast majority of the studies above were performed in 2D periodic boxes, with a CR current initialized but not sustained (\emph{undriven simulations}), while in astrophysical environments a parcel of fluid, e.g., in a shock precursor, is constantly exposed to ``fresh'' CRs. 
To mimic this more physical situation, \citet{reville+13} used a MHD+Vlasov code, where the CR distribution function is expanded in spherical harmonics satisfying a Fokker--Planck equation, to run \emph{driven simulations} in which the CR current is enforced to be constant.
In their simulations the maximum value of the magnetic field is always time-limited, and no final saturation of the magnetic field was reported.

\citet{kobzar+17} performed one full-PIC 2D simulation in a non-periodic box, which allows the authors to follow the spatio-temporal evolution of the instability; 
however, the relatively high current that they used led to the formation of a shock, and in general does not allow for the construction of a theory for arbitrary CR distributions.

In this paper, we study the Bell instability via hybrid simulations using the massively parallel code \dHy~\citep{gargate+07,haggerty+19a}, where ions are treated kinetically and their relativistic dynamics is retained. 
We perform {\it driven simulations} in which CRs are injected in the simulation box at a constant rate from the left side and are free to leave from the right, while the thermal background plasma and the electro-magnetic fields are subject to periodic boundary conditions.
This setup allows for a self-consistent coupling between CRs and thermal plasma, which eventually leads to the saturation of the instability. 
We explore a large range of parameters that characterize the CR current, always in the regime in which Bell is the fastest growing instability, and use a suite of 1D, 2D, and 3D simulations to investigate how the amplified magnetic field at saturation scales with the CR parameters.

We study the non-linear evolution of the driven Bell instability and provide the first simulation-validated prescription for the level of the final magnetic field  at saturation, i.e., that the final magnetic pressure is comparable to the initial net CR momentum flux.

The structure of the paper is as follows. 
In Section~\ref{sec:theory}, we briefly outline the possible CR distributions that may trigger the Bell instability and parametrize their free energy. 
Section~\ref{sec:simulations} describes our computational setup, the explored parameter space, and discusses the time evolution of a benchmark simulation.
In Section~\ref{sec:saturation}, we then describe the properties of the system at saturation and present the scaling of the total amplified magnetic field with the initial CR momentum flux.
Finally, in Section~\ref{sec:discussion} we discuss our results in the context of the current literature and outline some astrophysical implications of our findings before concluding in Section~\ref{sec:conclusions}.

\section{The free energy in streaming CRs}\label{sec:theory}
\label{sec/sims:setup}
\begin{figure}[t]
\centering
\includegraphics[width=\columnwidth]{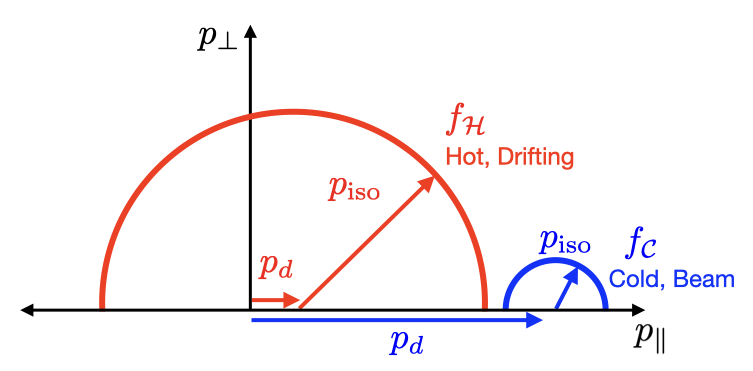} 
\caption{Schematic diagram of the initial CR distributions, distinguishing the hot and cold cases (also see Table \ref{tab:sims}).}
\label{fig:distro diag}
\end{figure}

Let us consider a population of CRs with isotropic monochromatic momentum $p\I \equiv \gamma\I m v\I$ in their rest frame, which drifts with velocity $\mathbf{v}\D =v\D \hat{\mathbf{e}}_x$ relative to the thermal plasma;
this corresponds to a current $\mathbf{J}\C = e n\C \mathbf{v}\D $, where $n_{\rm cr}$ represents the CR number density measured in the rest frame of the thermal plasma.

Let us fix the initial magnetic field $\mathbf{B} = B_0\hat{\mathbf{e}}_x$; 
its inclination with respect to $\mathbf{v}\D $ should not be important for the growth or the saturation of the Bell instability \citep[e.g.,][]{bell05,matthews+17}.
More specifically, the inclination may change the growth rate along the parallel and perpendicular direction, $k_{\rm \parallel}$ and $k_{\rm \perp}$ respectively, with respect to $\hat{\mathbf{e}}_x$, however the saturated magnetic field's amplitude may not change.

While the CR current and hence the growth rate only depend on $n\C$ and $v\D $, the net CR momentum and energy fluxes depend on $p\I$, too.
We consider two limiting cases: one in which the CRs are a \emph{cold beam} with $p\D \gg p\I$, and one in which they have a \emph{hot drifting} distribution with $p\D \ll p\I$, as illustrated in Figure \ref{fig:distro diag}.

\subsection{Bell's Prescription for Saturation}
According to the standard theory \citep{bell04,amato+09}, the right-handed, non-resonant modes grow faster than the resonant one if 
\begin{equation}\label{eq:kmaxrl}
     k_{\rm max} r_{\rm L} = \frac{n\C}{n_{\rm g}}\frac{v\D }{v_{A,0}}\frac{p_\perp}{mv_{A,0}} \equiv \xi_{\rm Bell} \gg 1 \; ,
\end{equation}
where $r_L(p_\perp)$ is the typical CR gyroradius. 

\citet{bell04} suggested that the instability saturates when the most unstable mode, calculated in the amplified field $\delta B$, becomes resonant with the CRs, i.e. $k_{\rm max}(\delta B) r_{\rm L}(\delta B)\sim 1$.
This is realized when 
\begin{equation}\label{eq:dBB}
     \left(\frac{\delta B}{B_0}\right)^2 \approx  \gamma\I\frac{n\C}{n_{\rm g}} \frac{ v\D c}{v_{A,0}^2} \equiv
     \frac{U\C}{U_{\rm B}}\frac{v\D }{2\,c} = \xi_{\rm Bell} \; ,
\end{equation} 
where $U\C\equiv\gamma\I m n\C c^2$ and $U_{\rm B}\equiv m n_{\rm g}v_A^2/2$ are the CR and magnetic energy densities, assuming that CRs are relativistic and $v\D$ is not; at saturation one then has $\delta B^2/B_0^2 \approx \xi_{\rm Bell}$.
\citet{blasi+15} came to a similar conclusion with an argument based on the dynamical balance between the magnetic tension and the $\mathbf{J}\C\times \mathbf{B}$ force. 

Note that Equation~\eqref{eq:dBB}: 1) is only \emph{similar} to the  ratio of CR and magnetic energy fluxes (the denominator is \emph{not} the magnetic energy flux, because waves do not move at $c$);
2) holds in the limit in which CRs are relativistic and the drift is not, since the $p_\perp$ that enters $r_{\rm L}$ is effectively $p\approx \gamma\I mc$;
3) has never been validated by means of self-consistent kinetic simulations.

For relativistic CRs, energy and momentum are essentially interchangeable, but for non-relativistic particles, or non-relativistic drifts, that is not the case.
Thus, it would be desirable to have a relativistically-covariant expression that expressed the field at saturation for an arbitrary CR distribution.

\subsection{A Covariant Formalism}
We start by considering the CR rest frame, where
the CR mass density is $\tilde{\rho}\C$, the total (including the rest mass) energy density is $e\C= \gamma\I \tilde{\rho}\C c^2$ and the isotropic pressure reads $P\C=\frac13 \gamma\I\tilde{\rho}\C c^2\beta\I^2$.
Then, we boost such a distribution into a frame that moves with velocity $\mathbf{v}_{\rm bst}$ and has a corresponding Lorentz factor $\Gamma_{\rm bst}=1/\sqrt{1-v_{\rm bst}^2/c^2}$.
In this frame, the CR density is $\rho\C=\Gamma_{\rm bst}\tilde{\rho}\C$ and the CR stress tensor reads \citep[see, e.g., \S133  of][]{landau6}:
\begin{equation}
T^{\alpha\beta}= (e\C+P\C) u\B^\alpha u\B^\beta + P\C\eta^{\alpha\beta},
\end{equation}
where $u\B^\alpha$ is the four-velocity constructed with $\mathbf{v}\B$  and $\eta^{\alpha\beta}$ is the Minkovski metric.
Explicitly, in components we have: 
$$T^{00} = \gamma\B^2 (e\C+P\C) - P\C \; ,$$
$$T^{0i} = T^{i0}=\gamma\B^2 (e\C+P\C) \frac{v\B^i}{c} \; ,$$
$$T^{ij} = \gamma\B^2 (e\C+P\C) \frac{v\B^iv\B^j}{c^2}+P\C\delta^{ij}$$
where $i,j=1,2,3$ and $\delta^{ij}$ is the Kronecker delta symbol. 
Here, $T^{00}$ has the usual meaning of energy density, while $T^{ij}$ is the flux along the $i$ direction of density of momentum $p_j$;
$T^{0i}$ does not have a non-relativistic counterpart, though $T^{0i}/c$ and $cT^{0i}$ are the density of momentum along $i $ and the energy flux along $i$, respectively. 

Note that, if CRs are relativistic, then the drift velocity $v\D$ that enters the CR current is generally \emph{smaller} than $v\B$ \citep[][]{gupta+21}, i.e.:
\begin{equation}
    v\D= \frac12 \int_{-1}^{1}{\rm d}\mu 
    \frac{\mu v\I+v\B}{1+\mu v\I v\B/c^2}\leq v\B,
\end{equation}
where $v_x=\mu|\mathbf{v}|$.
Since simulations are setup with an effective boost to the background plasma frame, we will provide the saturation as a function of $v\B$ rather than $v\D $.
Also, the average momentum along $x$ reads:
\begin{equation}
    \langle p_{{\rm cr},x}\rangle = \frac{\gamma\B}{2} \int_{-1}^{1}{\rm d}\mu (\mu p\I+\gamma\I m v\B)= \gamma\I p\B. 
\end{equation}

The questions that arise then are: 
1) can the saturation of the Bell instability be connected to the initial CR distribution?
2) which component(s) of $\mathbf{T}$ is the saturation of the Bell instability related to? 

In general, to drive the instability we need a finite current (i.e., charge flux), which also means a net momentum and energy flux.
We consider the following quantity:
\begin{equation}\label{eq:xi0}
\begin{split}
   \xi_0 & \equiv  \frac{T^{11}-P\C}{P_{B,0}} = 
    \frac{T^{01}}{P_{B,0}} \frac{v\B}{c} =
    \frac{T^{00}-\tilde{\rho}\C c^2}{P_{B,0}} \\
    & =  2 \gamma\I\gamma\B \frac{n\C}{n_{\rm g}} \frac{v\B^2}{v_{A,0}^2} \left( 1+\frac{1}{3}\frac{v\I^2}{c^2}\right),
\end{split}
\end{equation}
which expresses the \emph{net} CR momentum flux (consider the terms with $T^{11}$ and $T^{01}$), normalized to the magnetic pressure; 
the last equality derives from the definition of $\mathbf{T}$ and expresses the density in kinetic energy in the CR drift (term with $T^{00}$)\footnote{Technically, we expect CRs to isotropize in the wave frame moving with $v_A$ with respect to the background plasma.
Hence, the correct boost velocity to use in Equation~\eqref{eq:xi0} is the one that boosts the CR distribution to the wave frame rather than the background plasma frame.
This modification is necessary for CR distributions with $v\D\gtrsim v_A$, such as our cases $\mathcal{H}6-8$ in Table \ref{tab:sims}.}.
In the limit considered in Equation~\eqref{eq:dBB} (ultra-relativistic CRs and non-relativistic drift), from Equation~\eqref{eq:xi0} one has $\xi_0/\xi_{\rm Bell} \sim v\D/c\ll 1$, which suggests that a system may run out of free momentum before the field can saturate at equipartition of CR and magnetic fluxes.

\section{Hybrid Simulations}\label{sec:simulations}

We perform hybrid kinetic PIC simulations with \dHy, in which the ions are treated as macro-particles governed by the relativistic Lorentz force and the electrons are a massless, charge-neutralizing fluid \citep{haggerty+19a}. 
The parameters of the different runs are listed in 
Table \ref{tab:sims} and cover our benchmark run $\mathcal{B}$, several cold and hot CR distribution functions (see Figure \ref{fig:distro diag}, runs $\mathcal{C}$ and $\mathcal{H}$ respectively), as well as 1D and 3D versions of our benchmark (runs $\mathcal{D}$) and non-relativistic CRs (runs $\mathcal{N}$). 
Parameters are chosen in such  a way that Bell dominates over the resonant streaming instability ($\xi_0\gg 1$) and over the filamentary instability of \citet{niemiec+08} ($\gamma_{\max}\ll \omega_c$).

All simulations retain all three components of the velocity and fields.

\begin{table}[t]
	\centering
	\begin{tabular}{c c c c c c c}
		\hline
		{Run} &
		{$n\C/n_{\rm g}$} &
		{$p\B/(mv_{\rm A,0}) $} &
		{$p\I/(mv_{\rm A,0})$} &
		{$\xi_0$} \\
		\hline
		$\mathcal{B}$ & $10^{-3}$ &  $10^3$ & $1$ &  $195$ \\ 
		$\mathcal{B}2$ & $5 \times 10^{-4}$ &  $10^3$ & $1$ & $98$ \\ \hline
		$\mathcal{C}1$ & $10^{-3}$ & $1.5 \times 10^{3}$ & $1$ & $293$ \\
		$\mathcal{C}2$ & $10^{-3}$ & $5 \times 10^{2}$ & $1$ & $96$ \\
		$\mathcal{C}3$ & $10^{-3}$ & $2 \times 10^{3}$ & $1$ & $392$ \\
		$\mathcal{C}4$ & $10^{-3}$ & $7.5 \times 10^{2}$ & $1$ & $146$ \\
		$\mathcal{C}5$ & $10^{-3}$ & $10^{4}$ & $1$ & $1960$ \\
		$\mathcal{C}6$ & $10^{-3}$ & $5 \times 10^{3}$ & $1$ & $980$ \\
		$\mathcal{C}7$ & $10^{-3}$ & $3 \times 10^{3}$ & $1$ & $588$ \\
		$\mathcal{C}8$ & $10^{-3}$ & $8 \times 10^{3}$ & $1$ & $1568$ \\
		$\mathcal{C}9$ & $10^{-3}$ & $2 \times 10^{2}$ & $1$ & $35$ \\
		$\mathcal{C}10$ & $10^{-3}$ & $3 \times 10^{2}$ & $1$ & $56$ \\
		$\mathcal{C}11$ & $10^{-3}$ & $4 \times 10^{2}$ & $1$ & $76$ \\ \hline
		$\mathcal{H}1$  & $10^{-3}$ & $10^{3}$ & $10^{3}$ & $2601$ \\
		$\mathcal{H}2$  & $10^{-3}$ & $10^{3}$ & $500$ & $1312$ \\
		$\mathcal{H}3$  & $10^{-3}$ & $10^{3}$ & $10$ & $197$ \\
		$\mathcal{H}4$  & $10^{-3}$ & $10^{3}$ & $50$ & $233$ \\
		$\mathcal{H}5$  & $10^{-3}$ & $10^{3}$ & $10^2$ & $322$ \\ \hline
        $\mathcal{H}6$  & $9.9 \times 10^{-2}$ & $3$   & $10^3$ & $6$ \\
		$\mathcal{H}7$  & $3.6 \times 10^{-2}$ & $5$   & $10^3$ & $12$ \\
        $\mathcal{H}8$  & $2.4 \times 10^{-2}$ & $10$  & $500$  & $24$ \\
		$\mathcal{H}9$  & $6.8 \times 10^{-3}$ & $20$  & $10^3$ & $64$ \\
        $\mathcal{H}10$ & $1.4 \times 10^{-2}$ & $31$  & $400$  & $128$ \\
		$\mathcal{H}11$ & $8.2 \times 10^{-3}$ & $44$  & $500$  & $180$ \\
        $\mathcal{H}12$ & $2.2 \times 10^{-3}$ & $58$  & $400$  & $64$ \\
		$\mathcal{H}13$ & $4.2 \times 10^{-3}$ & $75$  & $10^3$ & $480$ \\
        $\mathcal{H}14$ & $3   \times 10^{-3}$ & $98$  & $10^3$ & $520$ \\
		$\mathcal{H}15$ & $1.5 \times 10^{-3}$ & $133$ & $10^3$ & $400$ \\
		$\mathcal{H}16$ & $           10^{-3}$ & $206$ & $10^3$ & $480$ \\ \hline
		$\mathcal{D}1$ & $10^{-3}$ & $10^{3}$ & $1$ & $195$ \\
		$\mathcal{D}3$ & $10^{-3}$ & $10^{3}$ & $1$ & $195$ \\ \hline
		$\mathcal{N}1$ & $5 \times 10^{-3}$ & $20$ & $1$ & $4$ \\
		$\mathcal{N}2$ & $5 \times 10^{-3}$ & $30$ & $1$ & $8$ \\
		$\mathcal{N}3$ & $5 \times 10^{-3}$ & $40$ & $1$ & $15$ \\
		\hline
	\end{tabular}
	\caption{List of the simulations used in the work, defined by: 
	the CR number density, $n\C$; 
	 the parallel CR momentum boost, $p\B $; 
	 the initial isotropic momentum, $p\I$;
	 and $\xi_0$, the CR net momentum flux (see Equation~\eqref{eq:xi0}). 
	The speed of light is $c=100 \; v_{\rm A,0}$ except for the $\mathcal{N}$-cases where it is $c=500 \; v_{\rm A,0}$. 
	Run $\mathcal{B}$ is our benchmark case, while runs $\mathcal{C}$ and $\mathcal{H}$ indicate cold-beam and hot-drifting cases, respectively; 
in all of these cases, boxes are 2D and measure at least $10^3\times 10^3 d_i^2$. 
	Finally, $\mathcal{D}1$ and $\mathcal{D}3$, correspond to our quasi-1D and 3D control runs and measure $10^3\times 50 ~d_i^2$ and $10^3\times 10^3\times 200~ d_i^3$, respectively.}
	\label{tab:sims}
\end{table}

Periodic boundary conditions are imposed for fields and thermal particles, while CRs experience periodic boundary conditions in the perpendicular ($\perp$) directions and open along the parallel ($||$) $x$-direction. 
This configuration mimics a system in which energetic CRs are being continuously injected, such as in a shock precursor \citep[][]{reville+13,caprioli+14b}. 

In most of the runs, the speed of light is $c=100\,v_{\rm A,0}$, except for the non-relativistic beam runs ($\mathcal{N}1-3$) where it is $c=500\,v_{\rm A,0}$.
The spatial resolution is two cells per ion skin depth and the time step is chosen to ensure that the Courant condition is satisfied. 
Since we limit ourselves to cases in which the growth rate is $\gamma\lesssim\Omega_{ci}$, then $k_{\rm max}d_i\lesssim 1$ and the fastest growing mode is always resolved. 

We initialize the simulation box with $4$ particles per cell for both background gas particles and CRs.
Finally, electrons are assumed to behave like a perfect fluid with adiabatic index of 5/3 \citep[see, e.g.,][]{caprioli+18}.

We checked the convergence of our results as a function of number of particles per cell, space/time resolution, and box size, finding that the minimum required box size must be larger than at least one CR gyroradius in the initial magnetic field.
Since it is notoriously difficult to deal with hot (i.e., subsonic) distributions close to open boundaries, for hot CR distributions we consider much larger computational boxes and perform our saturation analysis only in the center of the box, effectively avoiding any boundary effect.

\begin{figure*}[t]
\centering
\includegraphics[width=\textwidth]{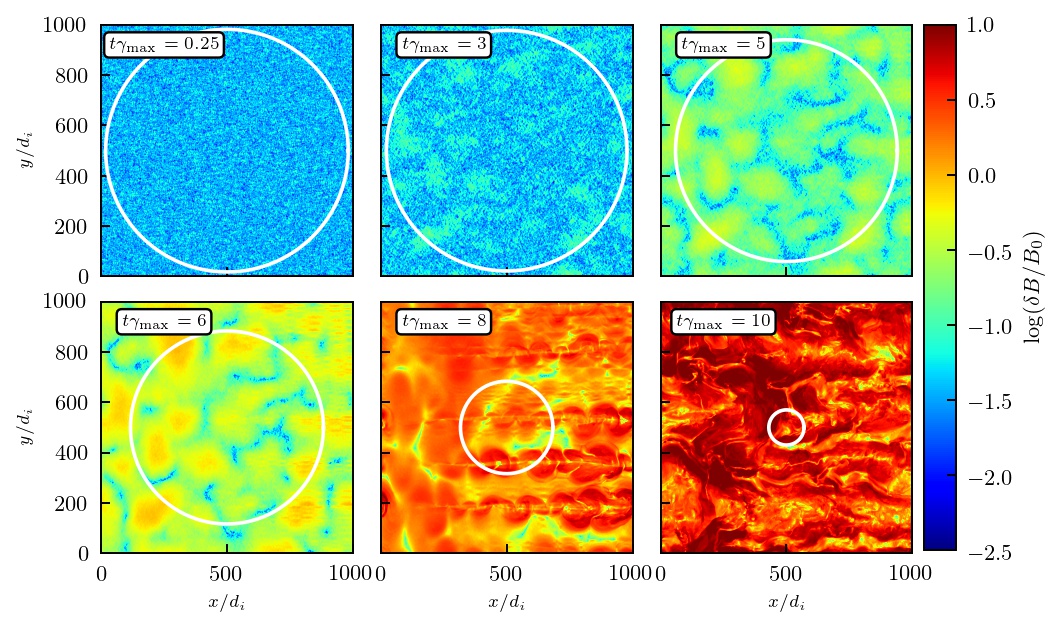}	
\caption{Time evolution of the $\log_{10}$ of the perpendicular component of the magnetic field, $\delta B = (B_y^2 + B_z^2)^{1/2}$, for our benchmark simulation (run $\mathcal{B}$ in Table~\ref{tab:sims}). 
Time is normalized to $\gamma_{\max}^{-1}$ and the color scale indicates the strength of the field in units of $B_0$.
The CR Larmor radius, defined in the amplified magnetic field, is represented by the white circles.
}
\label{fig:B-field evolution in sim box}
\end{figure*}

\subsection{The Benchmark Run}\label{sec:benchmark}
We now discuss the benchmark run ($\mathcal{B}$ in Table \ref{tab:sims}) to outline the features that are common to all of the simulations.
\begin{figure}[t]
\centering
\includegraphics[width=\columnwidth]{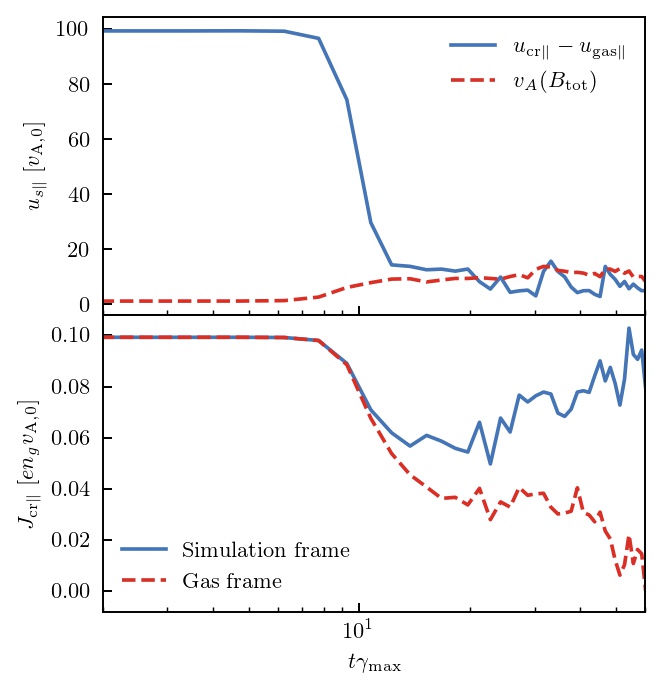} 
\caption{Top panel: Time evolution of the difference between the CR and gas bulk velocities for the benchmark run $\mathcal{B}$ in Table \ref{tab:sims}, compared with the Alfv\'en speed in the total magnetic field. 
Once $\delta B / B_0 \gtrsim 1$, CRs are scattered and their bulk motion drops abruptly, while momentum is transferred to the gas: the effective CR drift speed drops and at saturation becomes comparable to $v_A (B_{\rm tot})$, which attests to an effective coupling between CRs and thermal gas and the achievement of marginal stability. 
Bottom panel: CR current in both the simulation and the gas frame.}
\label{fig:bulk motion}
\end{figure}
As expected, CRs drive magnetic field perturbations that grow exponentially on a timescale $\sim \gamma_{\rm max}^{-1}$
\citep[e.g.,][]{riquelme+09,gargate+10,haggerty+19p}.
Figure~\ref{fig:B-field evolution in sim box} shows how magnetic structures grow in size and intensity.
Initially, such structures are much smaller than the CR gyroradius, represented by a white circle;  
both growth rate and maximally unstable wavelength are consistent with the linear theory predictions.
After a few to ten growth times, they become comparable to the CR gyroradii calculated in the amplified magnetic field $\gtrsim 10\,B_0$ (last panels).

It is interesting to look at the evolution of the first and second moments (bulk velocity and pressure) of the CR and gas distributions, indicated with $u_s$ and $P_s$, with $s={\rm CR, \; gas}$, respectively.
When $\delta B / B_0$ exceeds unity ($t\sim 5\gamma_{\rm max}^{-1}$), CRs start scattering off the magnetic perturbations and transfer momentum to the background gas. As a result, we see an initial abrupt drop in the CR bulk velocity, while the gas is set in motion in the direction of the CR drift, as discussed, e.g., by \cite{weidl+19a,weidl+19b}. 
At saturation CRs and gas are well coupled, and their relative speed reduces, as shown in Figure~\ref{fig:bulk motion}. 
More specifically, the difference between the two bulk velocities becomes comparable to the Alfv\'en speed in the total magnetic field at saturation (top panel of Figure \ref{fig:bulk motion}), which means that marginal stability is achieved. 
Since the CR current is driven, after the initial drop due to scattering, $J_{{\rm cr}\|}$ returns to its initial value in the simulation frame, but it keeps reducing in the gas frame (bottom panel of Figure~\ref{fig:bulk motion}), which slows the growth rate down.

\begin{figure}
\centering
\includegraphics[width=\columnwidth]{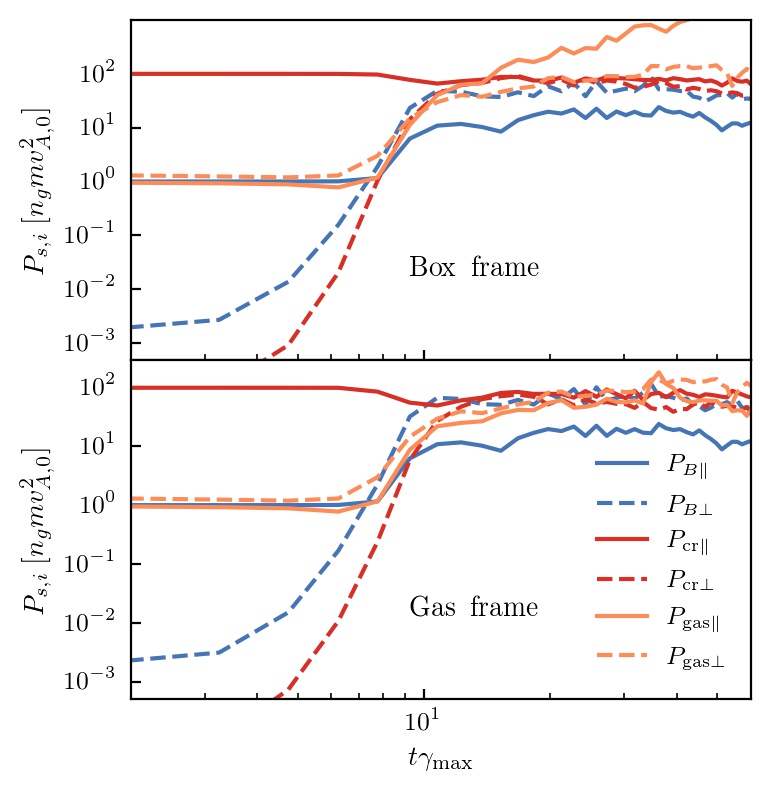}	
\caption{Evolution of the pressures of CRs, thermal gas, and magnetic field for our benchmark run $\mathcal{B}$, as seen in the simulation (top panel) or gas rest frame (bottom panel).
In both panels, $i$ denotes the direction (parallel or perpendicular) and $s$ denotes the species (CRs, magnetic field or gas).
\label{fig:abs and rel pressures}}
\end{figure}   

This picture is supported by Figure~\ref{fig:abs and rel pressures}, which shows how the thermal, CR and perpendicular (i.e., self-generated) magnetic pressures evolve in time; 
pressures are calculated both in the simulation and in the thermal gas frame (top and bottom panels of the figure).

The magnetic field initially grows exponentially until $t\sim 7 \; \gamma_{\rm max}^{-1}$, which marks the beginning of the ``secular'' stage of the instability, when $B$ keeps growing but with a slower growth rate. 
During this stage there is a rapid increase in the isotropic gas pressure; 
both $P_{\rm gas\perp}$ and  $P_{\rm gas\|}$ grow until $t\sim 10\; \gamma_{\rm max}^{-1}$, when the parallel momentum flux starts becoming larger than the transverse gas pressure.
The transferring of momentum from the CRs to the gas can be seen also in the slight drop of the parallel CR pressure, $P_{\rm cr\|}$, at $t \sim 10 \; \gamma_{\max}^{-1}$.

\begin{figure}[t]
\centering
\includegraphics[width=\columnwidth,clip=true, trim= 1 0 0 5]{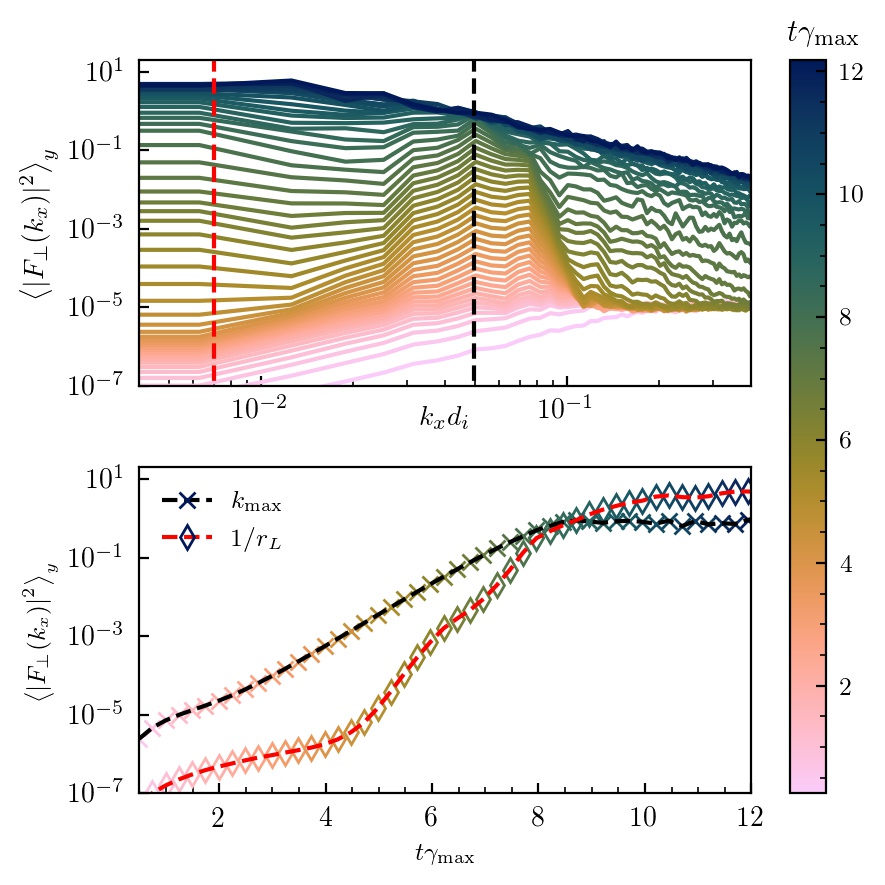}
\caption{Top panel: Transversely-averaged perpendicular magnetic power spectrum $\left< |F_\perp|^2\right>_y$, where $|F_\perp(k_x, y)|^2 = |{\mathcal{F}}[B_y(x,y)]|^2 + |{\mathcal{F}}[B_z(x,y)]|^2$ and $\mathcal{F}$ is the Fourier transform along the $x$ direction.
The power spectrum is plotted as a function of parallel wave number ($k_xd_i$) for the benchmark run $\mathcal{B}$, as it evolves in time (color coded).
The black dashed line shows the predicted fastest growing mode (Equation~\eqref{eqth:k_max Bell}), while the red dashed line corresponds to the mode in resonance with the CR gyroradius in the saturated magnetic field.
Lower panel: Time evolution of the power in the $k_{\rm max}$ mode (black dashed line, crosses) and in the resonant mode (red dashed line, diamonds). While the Bell mode grows much faster, at saturation most of the power is in quasi-resonant modes.}
\label{fig:Bperp_spec}
\end{figure}

This effect can be seen in more detail in Figure~\ref{fig:Bperp_spec}, which shows the temporal evolution of the perpendicular ($y,z$) magnetic power spectrum as a function of parallel wave number ($k_xd_i$, top panel) and time (bottom panel). 
The maximum growth is consistent with the predictions, with the black dashed line marking the location of $k_{\rm max}$ (Equation~\eqref{eqth:k_max Bell}).
The power in these modes grows over the exponential stage of the instability, while in the secular phase it is the power in longer wavelengths that takes over.
Such longer wavelength modes may be as large as the CR gyroradius in the saturated magnetic field (red dashed line, $1/r_{\rm L}$).
The power in these modes eventually surpasses that in Bell modes by the time the simulation reaches saturation, as shown in the bottom panel of Figure~\ref{fig:Bperp_spec}.

At $t\gtrsim 10 \; \gamma_{\rm max}^{-1}$,  the system reaches a state of pressure equilibrium where $P_{\rm gas \perp} \sim P_{\rm B\perp} \sim P_{\rm cr \|}$.           
The bottom panel of Figure \ref{fig:abs and rel pressures} illustrates this effect more clearly by showing the pressures in the frame moving with the bulk velocity of the plasma, $u_{{\rm gas}\|}$.
We note that the magnetic and thermal pressures grow together, suggesting that a fraction of the magnetic energy is dissipated into the background plasma, as already reported, e.g., by \citet{bell04,ohira+09b,gargate+10}.


\section{Magnetic Field at Saturation}\label{sec:saturation}
To understand how the saturation of the non-resonant instability depends on the CR distribution, we have preformed simulations that cover a wide parameter space (Table \ref{tab:sims}). 
In each simulation, parameters are selected so that we are in a regime where the Bell instability grows and dominates over any other instability. 
We chose to not vary the thermal speed of the background plasma, as we have found that ``WICE''-like effects can be important \citep[][]{reville+08a,zweibel+10, zacharegkas+19p,zacharegkas+21p}; 
we defer the study of this regime to a future work, stressing how the results presented here should apply to cases with plasma $\beta\equiv P_{\rm gas}/P_{\rm B} \ll 10$.
Similarly, a large CR current ($J\C > en_{\rm g}v_{\rm A,0}$) may trigger two-stream, Bunemann, or filamentation instabilities \citep{bret09}, a regime that we defer to future works as well.

\subsection{A simulation-validated prescription for magnetic fields at saturation}

\begin{figure}
	\centering
\includegraphics[width=\columnwidth, clip=true,trim= 3 0 1 0
]{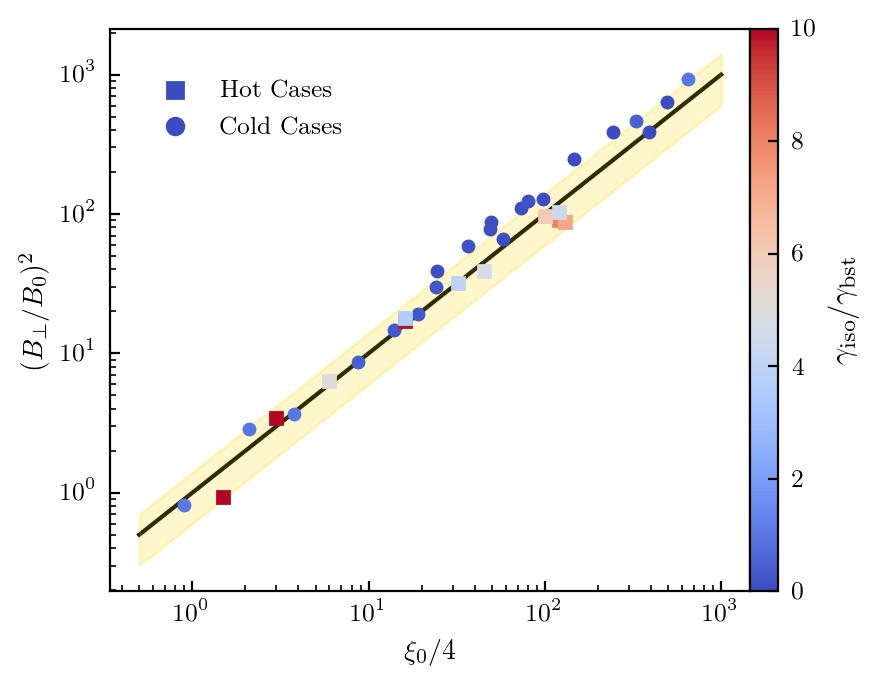}
	\caption{\label{fig:saturation_plot} 
	Transverse magnetic field at saturation, as a function of the initial CR anisotropic momentum flux normalized to the initial magnetic pressure, $\xi_0$ (Equation~\eqref{eq:xi0}). 
	The line corresponds to $(B_\perp/B_0)^2= \xi_0/4$. Each point represents one case from Table \ref{tab:sims}, excluding the cases $\mathcal{D}1$ and $\mathcal{D}3$ because of their different dimensionality. 
    The gold band shades the $40\%$ area around the line.}
\end{figure}

The amplitude of the transverse magnetic field at saturation is shown in Figure \ref{fig:saturation_plot} for many hot/cold runs (color coded), as a function of the normalized net CR momentum flux defined in Equation~\eqref{eq:xi0}. 
The black line corresponds to 
\begin{equation}\label{eq:saturation}
    \frac{B_\perp^2}{B_0^2}= \frac{\xi_0}{4}
    \Leftrightarrow P_{B,\perp}= \frac{\xi_0}{4} P_{B,0},
\end{equation} 
indicating that $\xi_0$ is a good parameter for estimating the total magnetic field amplification produced by generic CR distributions.
This result should apply in all the cases in which the Bell modes grow faster than the resonant cyclotron instability ($k_{\max} r_{\rm L} \simeq \xi_{\rm Bell}\gg 1$).
Also, it may apply even in the presence of instabilities (e.g., Weibel, two-stream, etc.) that grow faster than Bell, but at scales small enough that when their fastest-growing modes saturate, the CR current is not strongly affected and still able to drive Bell modes.
This would be similar to what happens for the Bell instability, in that at saturation most of the magnetic power is never on scales comparable to $k_{\max}$, but rather on scales of the order of the CR gyroradius, as shown in Figure~\ref{fig:Bperp_spec}.

As our new prescription shows, the relevant quantity is the CR momentum flux $\xi_0$ rather than the CR current itself, which means that saturation does not necessarily correlate with growth rate.
One non-trivial implication of this finding is that the same electric current made of more energetic CRs should result in a larger magnetic field amplification, which may be relevant for CRs escaping their sources \citep[e.g.,][]{cristofari+21,schroer+21,schroer+22}.
We tested this expectation explicitly with our simulations $\mathcal{C}$2, $\mathcal{B}$ and $\mathcal{C}$7, which share the same CR current but saturate at different values of $\delta B$ correlating with $\xi_0$.

The amplification described by Equation~\eqref{eq:saturation} can be explained with the following simple argument based on the bottom panel of Figure \ref{fig:abs and rel pressures}.
At saturation the system achieves equipartition between: 
the 3 components of the thermal plasma pressure, the 3 components of the CR pressure, and the 2 transverse components of the amplified magnetic field. 
Since the initial free momentum is $\xi_0$, and since 2 out of the 8 final channels are in $P_{\rm B\perp}$, Equation~\eqref{eq:saturation} naturally ensues. 

In principle, another channel in which the injected energy may go is the bulk motion; 
however, before saturation ($t\lesssim 10 \gamma_{\max}^{-1}$) the thermal gas is not sped up much, while after saturation all of the energy goes into accelerating the coupled system of CRs+thermal plasma+magnetic fields (top panel of Figure \ref{fig:abs and rel pressures}).
Thus, the bulk motion is not an effective degree of freedom of the system, but rather the ultimate sink of all the energy injected after saturation.

Finally, we note that the effective CR drift becomes comparable with the Alfv\'en speed, eventually, but this happens \emph{after} the magnetic field amplification has stopped because the system ran out of free momentum;
the achievement of marginal stability can therefore be interpreted as a manifestation of the efficient coupling between CRs and magnetic fields \citep[e.g.,][]{zweibel17}.


\section{Discussion}\label{sec:discussion}
\begin{figure}
    \centering
    \includegraphics[width=1\linewidth]{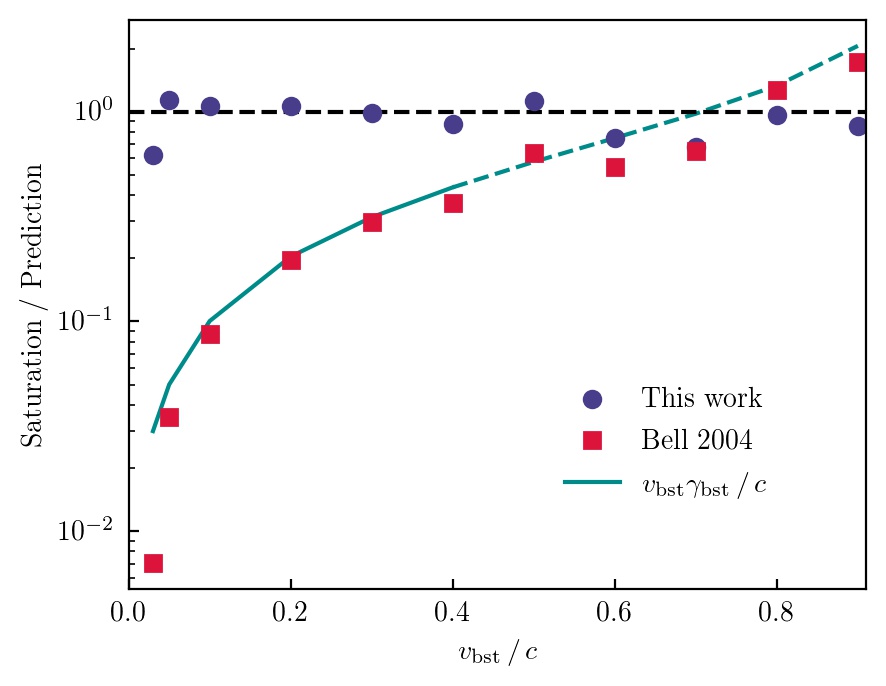}
    \caption{Saturated value of $(\delta B_\perp/B_0)^2$ measured in simulations over the one predicted by either Equation~\eqref{eq:dBB} \citep[][]{bell04} or Equation~\eqref{eq:saturation}; 
    they differ by a factor of $v\B\gamma\B/c$, which means that Bell's prescription overestimates the measured values when the boost is non-relativistic. We show the cases $\mathcal{H}6-16$ as these sample the $v\B$-parameter space.}
    \label{fig:comparison}
\end{figure}

Arguably the main application of the non-resonant instability is for DSA in SNR shocks, as highlighted already by \cite{bell04}, who derived Equation~\eqref{eq:dBB} in the limit of the non-relativistic drift of ultra-relativistic CRs with a power-law distribution in momentum $\propto p^{-4}$.
In the same limit, Equation~\eqref{eq:xi0} would return a perpendicular amplified magnetic field
\begin{equation}
     \left(\frac{\delta B_\perp}{B_0}\right)^2 = \frac{1}{4}\xi_0 \approx  \frac{2}{3}\gamma\I\frac{n\C}{n_{\rm g}} \frac{ v\D^2}{v_{A,0}^2},
\end{equation}
where we used $v_{\rm bst}\approx v\D$ for non-relativistic boosts.

This field is much smaller than the one provided by Equation~\eqref{eq:dBB}, since $v\D\sim v_{\rm shock}\ll c $ for SNRs.
Figure \ref{fig:comparison} shows the ratio of $(\delta B_\perp/B_0)^2$ at saturation in simulations compared with either Bell's prescription (Equation~\eqref{eq:dBB}) or the new one (Equation~\eqref{eq:xi0}).
Note that in Figure \ref{fig:comparison} we multiplied Bell's prescription by a factor $2/3$ in order to get an estimate of $\delta B_\perp^2$ rather than the total $\delta B^2$, assuming isotropic turbulence.

The current estimates of the maximum energy achievable in SNRs, which rely on Bell's saturation \citep[e.g.,][]{reville+12,bell+13}, suggest that typical SNRs can accelerate particles up to maximum rigidities of fractions of a PV, a factor of a few to ten smaller than the rigidity of the CR knee.
More precisely, the problem is not to produce PeV particles, but to produce \emph{enough} of them such that the overall CRs spectrum rolls over at the knee \citep[e.g.,][]{cardillo+15,cristofari+21,cristofari+22,diesing+21}.

While our saturation prescription (Equation~\eqref{eq:saturation}) seems to exacerbate this issue, thus challenging the idea that SNRs can be the sources of Galactic CRs, one has to remember that ahead of the shock there is a flux of escaping particles \citep[e.g.,][]{zirakashvili+08,caprioli+09a,caprioli+10c,bell+13,caprioli+14b}, which have momentum close to $p_{\rm max}$, i.e., the instantaneous maximum momentum. 
Such particles have a very anisotropic distribution that can be effectively described as a cold beam with $p\B\approx p_{\rm max}$ and $v\B\approx c$, but with number density $n_{\rm esc} \simeq \frac{v_{\rm sh}}{c} n\C$, with $n\C$ measured at the shock \citep[at least for a $p^{-4}$ CR spectrum, see e.g.,][]{caprioli+09a,bell+13}. 
Hence, assuming isotropic turbulence, the total field, i.e. $\delta B^2=\frac{3}{2}\delta B_\perp^2$, produced by escaping particles reads
\begin{equation}
     \left(\frac{\delta B}{B_0}\right)^2 \approx  
    \frac{3}{4}\frac{n_{\rm esc}}{n_{\rm g}} \frac{c \,p_{\rm max}}{m v_{A,0}^2} \approx
    \frac{3}{4} \frac{n\C}{n_{\rm g}} \frac{v_{\rm sh}p_{\rm max}}{mv_{A,0}^2},  
\end{equation} 
which is exactly Bell's prescription (Equation~\eqref{eq:dBB}, modulo a factor of order unity).
Eventually, our results do not quantitatively change the estimates of the maximum energy achievable in SNRs \citep[e.g.,][]{reville+12,bell+13,cardillo+15,cristofari+21,cristofari+22,diesing+21}, though suggest that most of the magnetic field amplification upstream of SNR shocks must be driven by the current in escaping particles, rather than by the current due to the anisotropy of CRs diffusing in the precursor. 
This also means that if the CR spectrum is steeper than $p^{-4}$ the saturated magnetic field may be smaller than the standard prediction, and thus lead to smaller maximum energies, as pointed out, e.g., by \cite{cristofari+22}.

\section{Conclusions}\label{sec:conclusions}
We have used controlled hybrid simulations to investigate the saturation of the Bell instability \citep{bell04,bell05} for a wide range of CR distributions, spanning from cold beams to hot-drifting cases (Figure \ref{fig:distro diag} and Table \ref{tab:sims}).
We used a suite of 1D, 2D, and 3D runs to assess the final amplitude of the self-generated magnetic field and our main findings can be summarized as follows:

\begin{itemize}
    \item During the linear stage of the instability the CR current is undisturbed by the small-wavelength waves that are excited at $k_{\rm max}$; when $\delta B/B_0\gtrsim 1$ CRs start scattering and the amplitude of the magnetic field grows linearly rather than exponentially. 
    \item After about 10 growth times, the pressure in CRs, magnetic fields, and thermal plasma become comparable (Figure \ref{fig:abs and rel pressures}) and the instability saturates.
     \item At saturation CRs and thermal plasma are well coupled: the effective drift speed becomes comparable to the Alfv\'en speed in the amplified field and marginal stability is achieved (Figure \ref{fig:bulk motion}).
     \item In the asymptotic state most of the magnetic power is not at the fastest growing mode, but rather at larger scales, comparable to the CR gyro-radius in the amplified field (see Figures \ref{fig:B-field evolution in sim box} and \ref{fig:Bperp_spec}).
    \item While the growth rate depends on the CR current, the total magnetic field amplification that can be achieved depends on  the momentum flux in CRs, generally parametrized by Equation~\eqref{eq:xi0}. 
    In particular, simulations suggest that the final pressure in transverse magnetic fields is $\sim 1/4$ of the initial CR pressure flux (Equation~\eqref{eq:saturation} and Figure \ref{fig:saturation_plot}). 

    \item The simulation-validated prescription (Equation~\eqref{eq:saturation}) for the saturated magnetic fields yields amplification factors \emph{smaller} than the original prescription by Bell (Equation~\eqref{eq:dBB}), especially for non-relativistic boosts (Figure \ref{fig:comparison}). 

    \item The results above do not change the estimate of the maximum energy achievable in SNRs, where most of the field amplification is driven by the anisotropic beam of ultra-relativistic particles escaping upstream, rather than by CR diffusing in the shock precursor.    
\end{itemize}

These results provide the first prescription, validated by self-consistent kinetic simulations, that can be used to estimate the total magnetic field amplification ensuing from an arbitrary distribution of anisotropic CRs, as long as electron physics is not important \citep[i.e., that the current is not too strong to drive instabilities other than Bell's, see][]{bret09,niemiec+08,riquelme+09} and the thermal plasma is relatively cold \citep[e.g.,][]{reville+08a,zweibel+10}.
We defer the study of saturation in such systems to future works. 

Quantifying the level of magnetic field amplification in SNRs is important not only for estimating the maximum energy achievable via DSA, but also the dynamics of the shock, the overall shock compression, and the slope of the accelerated particles \citep[][]{haggerty+20,caprioli+20,diesing+21, cristofari+22}.

\begin{acknowledgements}
We thank the Reviewer for their constructive comments. 
Simulations were performed on computational resources provided by the University of Chicago Research Computing Center and on  TACC's Stampede2 thourgh ACCESS (formally XSEDE) allocation (TG-AST180008).
We wholeheartedly thank Ellen Zweibel, Pasquale Blasi, Elena Amato and Lorenzo Sironi for interesting and stimulating discussions. 
This work of D.C. was partially supported by NASA through grants 80NSSC20K1273 and 80NSSC18K1218 and NSF through grants AST-1909778, PHY-2010240, and AST-2009326, while the work of C.C.H was supported by NSF FDSS grant AGS-1936393 and NASA grant 80NSSC20K1273.

\end{acknowledgements}

\appendix

\section{Effects of simulation dimensionality}

\begin{figure}[t]
\centering
\includegraphics[width=0.65\columnwidth,clip=true, trim= 1 5 10 5]{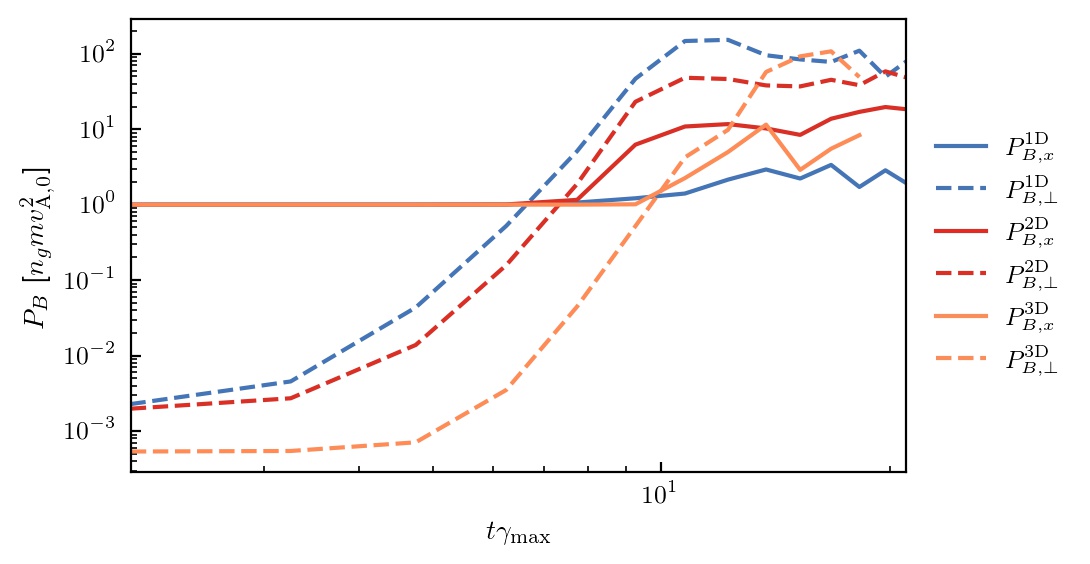}
\caption{Time evolution of the parallel (solid) and perpendicular (dashed and dotted) components of the magnetic field for the benchmark run, in 1D, 2D and 3D configurations (runs $\mathcal{D}1$, $\mathcal{B}$, $\mathcal{D}3$); 
note that run $\mathcal{D}_1$ has $50d_i$ in the $y$ direction, so it is not exactly 1D, which allows for a $B_x$ that is not strictly constant.
The $B_z$ component in 2D is a good proxy for both transverse components in full 3D setups.}
\label{fig:1D vs 2D vs 3D}
\end{figure}

In 2D hybrid simulations, we report an anisotropy in the transverse (self-generated) magnetic field. 
Specifically, for our 2D boxes in the $x-y$ plane, we find that typically $B_z \gtrsim B_y$ and that the ratio of such transverse components may reach a factor of a few, depending on the simulation parameters. 
This makes it hard to choose which component of the magnetic field must be used to calculate the saturated value of the field. 
In quasi-1D boxes, i.e., when $L_y \ll L_x$ and $L_z=0$, we find that at saturation $B_y \sim B_z > B_x$, and that $B_x \sim B_0$ as expected. 
In 3D simulations, instead, we find that $B_y \sim B_z$, similarly to the 1D case, and that the two components are comparable to the $B_z$ one in the 2D box. 
These scalings are shown in Figure~\ref{fig:1D vs 2D vs 3D} where we compare simulations $\mathcal{D}1$, $\mathcal{D}3$ and $\mathcal{B}$. 
The former two are similar to our benchmark simulation $\mathcal{B}$, but with 1D and 3D boxes, respectively.

This test indicates that the $B_z$ component in 2D simulations might be a good proxy for the behavior of both perpendicular components in more realistic 3D setups. 
However, the difference between $\delta B_\perp= (B_y^2+B_z^2)^{1/2}$ and the proxy $\sqrt{2}B_z$ is only about $20-30\,\%$.
Therefore, we use $\delta B_\perp$ to quantify the field at saturation in this paper keeping in mind that this might be slightly underestimating the actual value in full 3D.
Such an uncertainty is also conveyed by the yellow envelop around the theoretical prescription (Equation \ref{eq:xi0}) in Figure \ref{fig:saturation_plot}.

\bibliography{Total}

\end{document}